\documentclass[twocolumn,amsmath,amssymb,aps,superscriptaddress,prd,10pt,showpacs,nofootinbib,floatfix]{revtex4-2}

\usepackage{hyperref,amsmath,amssymb,wasysym,units,pifont,footnote,booktabs,inputenc, ulem}

\usepackage[dvipsnames]{xcolor}
\usepackage{graphicx}
\usepackage{lineno}
\usepackage{upgreek}
\usepackage{braket}
\usepackage{float}

\begin{document}

\title[Mechanical Characterisation of Silicon for the ETpathfinder Test Masses]{Mechanical Characterisation of Silicon for the ETpathfinder Test Masses}

\author{Guido Alex Iandolo}
\email{g.iandolo@maastrichtuniversity.nl}
\author{Alex Amato}
\affiliation{Maastricht University, P.O. Box 616, 6200 MD Maastricht, The Netherlands}
\affiliation{Nikhef, Science Park 105, 1098 XG Amsterdam, The Netherlands}
\author{Gianpietro Cagnoli}
\affiliation{Institut Lumière Matière, UMR5306 CNRS Université Claude Bernard Lyon 1,
Villeurbanne, France}
\author{Alessandro Delmonte}
\affiliation{Maastricht University, P.O. Box 616, 6200 MD Maastricht, The Netherlands}
\author{Jan-Simon Hennig}
\affiliation{Maastricht University, P.O. Box 616, 6200 MD Maastricht, The Netherlands}
\affiliation{Nikhef, Science Park 105, 1098 XG Amsterdam, The Netherlands}
\affiliation{Fraunhofer Centre for Applied Photonics, Technology and Innovation Centre, Glasgow, G1 1RD, United Kingdom}
\author{Margot Hennig}
\affiliation{Maastricht University, P.O. Box 616, 6200 MD Maastricht, The Netherlands}
\affiliation{Nikhef, Science Park 105, 1098 XG Amsterdam, The Netherlands}
\affiliation{SUPA, School of Physics and Astronomy, University of Glasgow, Glasgow, G12 8QQ, Scotland}
\author{Sebastian Steinlechner}
\author{Janis W{\"o}hler}
\author{Stefan Hild}
\affiliation{Maastricht University, P.O. Box 616, 6200 MD Maastricht, The Netherlands}
\affiliation{Nikhef, Science Park 105, 1098 XG Amsterdam, The Netherlands}
\author{Jessica Steinlechner}
\email{jessica.steinlechner@ligo.org}
\affiliation{Maastricht University, P.O. Box 616, 6200 MD Maastricht, The Netherlands}
\affiliation{Nikhef, Science Park 105, 1098 XG Amsterdam, The Netherlands}
\affiliation{SUPA, School of Physics and Astronomy, University of Glasgow, Glasgow, G12 8QQ, Scotland}
\date{\today}

\begin{abstract}

The next generation of gravitational-wave detectors, such as the Einstein Telescope, is designed to reduce noise in a wide band of frequencies compared to the current generation, through the use of new technologies. ETpathfinder, designed as an R\&D facility for these technologies, is a prototype for which the mirrors were chosen to be made of crystalline silicon, produced by the Leibniz-Institut f\"{u}r Kristallz\"{u}chtung. This material choice was made to pave the way for a low thermal noise level at cryogenic temperatures in the Einstein Telescope. This paper shows the mechanical loss of silicon designated to become the test masses for ETpathfinder in the range between room temperature and 53\,K. In addition, the effect of the anisotropic nature of silicon on the measurement procedure is addressed. Predictions are made of the contribution of the mirror substrate material to the overall ETpathfinder noise budget.

\end{abstract}

\maketitle

\section{Introduction}

The detection of gravitational waves began in 2015~\cite{PhysRevLett.116.061102}, opened a new window into the Universe and established the basis for a new type of multi-messenger astronomy~\cite{GW170817, Multi-messenger_Observations}. The well-known behavior of Michelson interferometers enabled the construction of kilometer-scale detectors that employ heavy mirrors as test masses (TMs) of the gravitational field to detect gravitational waves: the LIGO Livingston and LIGO Handford detectors in the United States and the Virgo detector in Italy. Correspondingly approved in 1990 (LIGO) and 1993 (Virgo) and inaugurated in 1999 and 2000~\cite{universe2030022}, it took 15 years of improvements to start detecting the first gravitational-wave signals with the corresponding advanced configurations~\cite{Aasi_2015,Acernese_2015}. Later, the KAGRA detector, located in Japan, started participating in observation runs alongside LIGO and Virgo~\cite{KAGRA_run_join}.

Although the current generation of gravitational-wave detectors is the result of engineering that extends beyond the established state of the art, the sources of noise that limit their sensitivity are multiple and are not always easy to identify or mitigate. 
The next generation of detectors is planned to observe phenomena in a wider frequency band through the implementation of new technologies and the improvement of existing ones. The Einstein Telescope~\cite{ETdesignstudy} will embody this vision, consisting of three detectors arranged in a triangular formation. Each detector will consist of two interferometers in a so-called xylophone configuration, one optimized for high frequencies and one for low frequencies.

One of the most dominant noise sources, in the most-sensitive frequency range, is thermal noise of the highly-reflective coatings of the interferometers' test mass mirrors. This noise source is material related, originating from the Brownian motion of the atoms and molecules the mirrors are comprised of. It can be derived from the measurement of the mechanical properties of the material itself and quantified by the so-called mechanical loss angle, as shown in~\cite{Saulson}. The thermal noise amplitude spectral density of the mirror coatings is inversely proportional to the square root of the frequency $f$ and to the square root of the mirror temperature $T$. Consequently, thermal noise is more relevant for the low-frequency detectors within the Einstein Telescope. Furthermore, it shows that a reduction of this type of noise can be achieved by cooling the mirrors. Not only coating thermal noise benefits from a temperature reduction, but also substrate and suspension thermal noise.

Current generation detectors operate at room temperature with mirror substrates made of fused silica due to its low mechanical loss, which leads to a low contribution of the mirror substrates to the detectors' thermal noise~\cite{10.1063/1.1394183}. However, at cryogenic temperatures, the contribution of the silica to the overall thermal noise increases significantly due to an increase in its mechanical loss by several orders of magnitude~\cite{schroeter2007mechanical, 10.1063/1.1721138, 1954JAP.25.402F}.
For this reason, new materials, particularly crystalline silicon and sapphire, are being considered as mirror-substrate material because of their low mechanical loss at cryogenic temperatures~\cite{R_Nawrodt_2008, UCHIYAMA19995, 10.1117/12.459019}.

ETpathfinder, a prototype detector in Maastricht, is designed to test key technologies needed for the Einstein Telescope, with a focus on cryogenics, featuring silicon test masses~\cite{Utina_2022}. Large, high-purity, undoped, ingots produced using the Float-Zone technique were supplied by the Leibniz-Institut für Kristallzüchtung (IKZ) in Berlin-Adlershof\footnote{\url{https://www.ikz-berlin.de/}}. The ingots were subsequently divided into sections intended to become test masses in ETpathfinder, along with surrounding witness wafers used for characterization purposes.

This article presents mechanical loss measurements of test samples along the length of the ingots. Furthermore, the radial loss distribution and the comparability between ingots are investigated.
Considerations regarding the anisotropic properties of silicon, which influence the mechanical loss measurement due to the presence of an imperfect nodal area at the center of the samples, are made to estimate the intrinsic material mechanical loss.
Finally, mechanical losses at both room temperature and cryogenic temperature are presented and predictions are made of the implications on the ETpathfinder sensitivity. 

\section{Sample preparation and nomenclature}
\label{Labelling and sample preparation}

\begin{figure}[b]
    \centering
    \includegraphics[width=8cm]{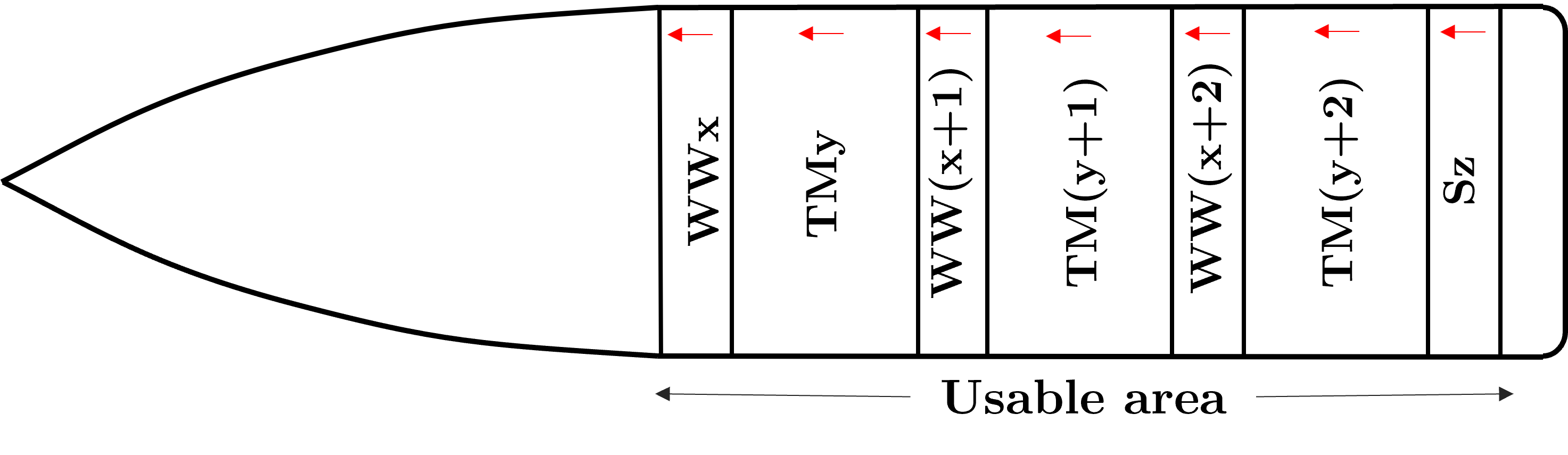}
     \caption{Schematic of a generic ingot, where the identification numbers increase independently. Arrows were etched into the side surface are represented in the figure with red arrows.}
    \label{fig:Ingotx}
\end{figure}

\begin{table}[t]
 \caption{Dimensions of samples for a generic ingot.}
    \label{tab:Dimension_ingot3}
\begin{ruledtabular}
\begin{tabular}{ccc}
Sample &Thickness (mm) & Diameter (mm) \\
\hline
Test Masses (TMs) & 82 & 152-155\\ 
Witness Wafers (WWs) & 32 & 152-155\\ 
Small wafers (S) & [18,21.5] & 152-155\\ 
\end{tabular}
\end{ruledtabular}
\end{table}

IKZ supplied four $\braket{100}$ oriented Float-Zone silicon ingots. Each ingot consists of three main zones: a non-cylindrical zone, a usable zone, and a final residual zone, containing the highest concentration of impurities -- see Fig.~\ref{fig:Ingotx} for a schematic of an ingot. 

Test mass candidates, TMs, for ETpathfinder, with a target diameter of $15$\,cm and a thickness of $8$\,cm were obtained from the usable zone of the ingots. In addition, $3.2$\,cm thick witness wafers, WWs, were cut from both sides of each TM -- see Fig.~\ref{fig:Ingotx}. The WWs surrounding each TM were intended for characterization purposes to estimate the properties of the TM material, without the need to directly characterize the TMs, avoiding possible damage and contamination. On the basis of the usable zone of each ingot and the fixed dimensions of the TMs and WWs, additional smaller wafers, labeled S, with a thickness depending on the residual usable area, were produced from some ingots. Table~\ref{tab:Dimension_ingot3} lists the nominal thickness and diameter of each sample. Arrows were etched into the side surface of each TM and WW to allow identifying the relative orientation during subsequent cutting and handling.

\begin{figure}[b]
    \centering
    \includegraphics[width=7.8cm]{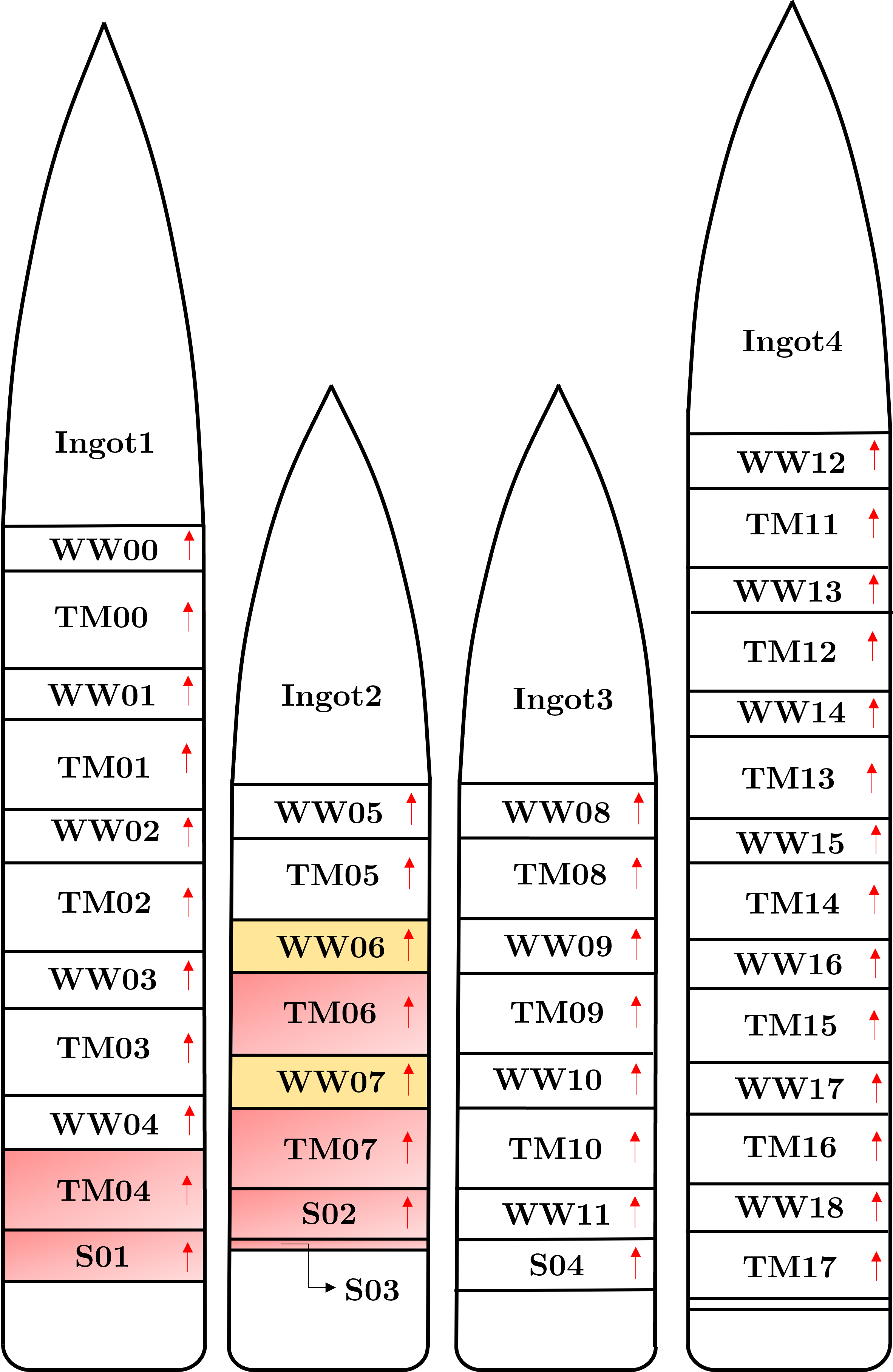}
     \caption{Schematics of the four silicon ingots. The proportions of the samples are not to scale and are shown for illustrative purposes only. Red shading is used to denote broken samples while yellow shading is used for partially broken samples.}
    \label{fig:All_ingots_marked}
\end{figure}

The four ingots were assigned identification numbers ranging from $1$ to $4$. Within each ingot, the WW and TM identification numbers are independent and increase along the ingot, as shown in Fig.~\ref{fig:Ingotx}. The numbering system starts from x = 00 in Ingot1 and does not reset between ingots. A comprehensive overview of the samples obtained from the four ingots is presented in Fig.~\ref{fig:All_ingots_marked}. Some samples broke during the cutting procedure. Broken samples are marked red and partially broken samples are marked yellow.

\begin{figure}[t]
    \centering
    \includegraphics[width=6cm]{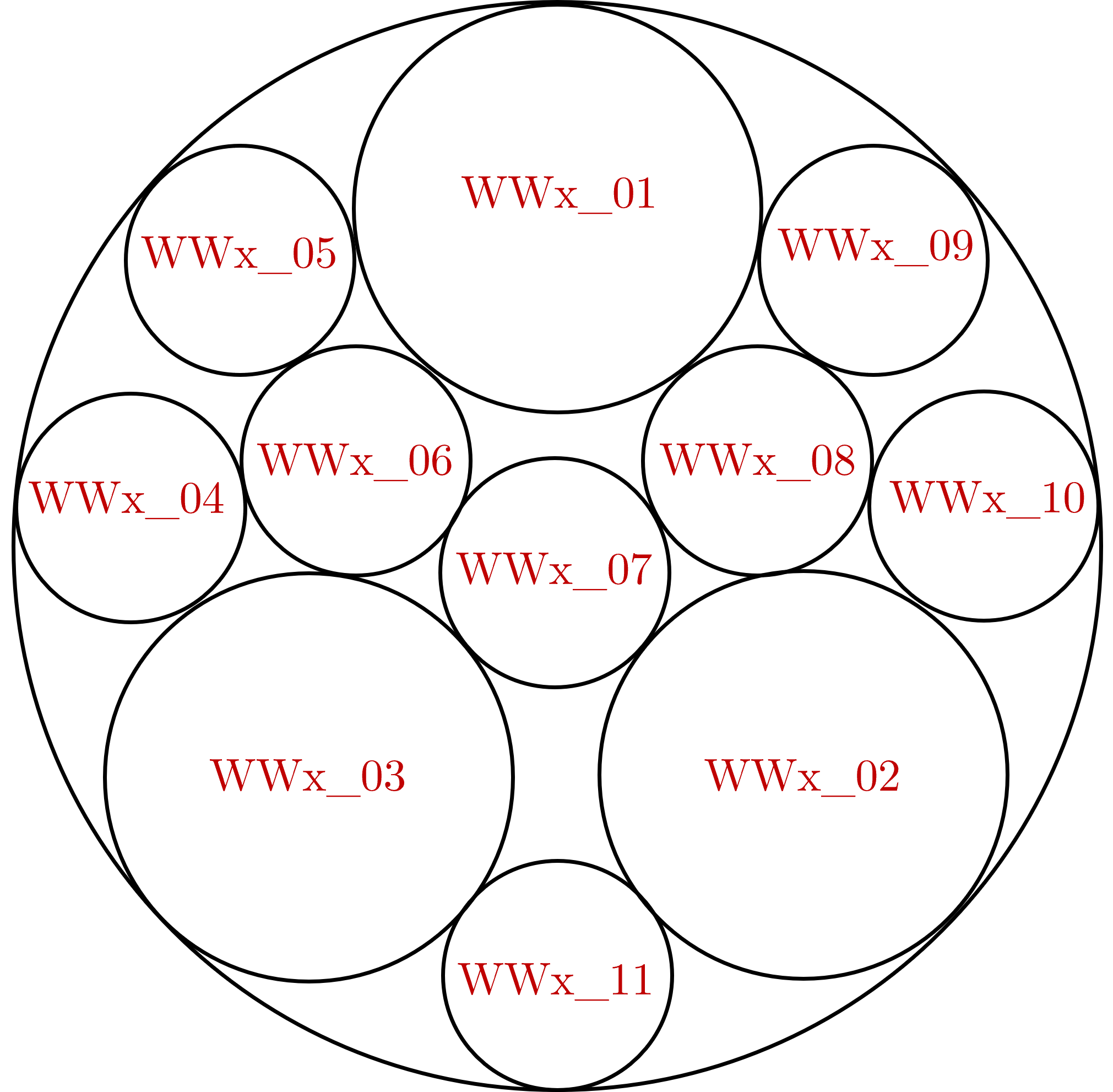}
    \caption{Layout and nomenclature of samples from a generic WW, showing the position of the 2" and 1" samples.}
    \label{wwcut}
\end{figure}

Each WW was subdivided into three 2" diameter samples and eight 1" diameter samples (see Fig.~\ref{wwcut}), due to its large size, which makes it incompatible with the dimensions of our mechanical loss measurement setup. The smaller samples are intended for measurements of optical properties, while the 2" samples were designed for mechanical-loss measurements. For the witness wafers investigated in this paper, the 2" samples were further sliced into thinner sections of $5$\,mm thickness and all surfaces were polished. The thickness of the 2" disks was selected to ensure that the thermoelastic loss peak occurs at frequencies lower than our measured frequency band, which is between 10 kHz - 100 kHz~\cite{Silenzi_2024}.

The cutting pattern shown in Fig.~\ref{wwcut} was chosen to maximize the number of samples obtained from each WW, while maintaining the position of a small-diameter sample in the center. This region represents the central area of a TM, where, inside an interferometer, the laser light passes through, making the optical properties particularly relevant.

Each sample was given an identification code consisting of two numbers: one for the originating WW and the other for the position within the WW. 

\section{Room temperature mechanical loss characterization}
\label{Room temperature mechanical loss characterization}

In this section, we present the results of mechanical loss measurements performed on representative $2"$ samples.
The mechanical losses of the WWs in their original, pre-cutting form were not measured. However, loss values for a large $\braket{100}$-oriented silicon sample (98 mm in diameter and 100 mm thick) can be found in~\cite{theses_murray}.

A gentle nodal suspension system (GeNS)~\cite{Cesarini_2009} was used, in which the mechanical loss was measured using the ring-down method. During the measurements, the sample was suspended in equilibrium on a silicon lens. Due to the cylindrical symmetry of the samples, the point of equilibrium, and therefore of contact, is found to be at a nodal point for resonance modes with radial nodal lines -- see  Sec.~\ref{Sec:Result_interpretation} for a more detailed discussion. This allows the mechanical loss of these modes to be measured as if the disk was subject to free vibration, thus excluding sources of elastic energy dissipation due to clamping. This operating principle, together with the fact that measurements are carried out under vacuum, ensures that the measured internal friction is exclusively that of the material.
If the cylindrical symmetry or the isotropy of the sample is broken, not all modes within the same family exhibit nodal lines passing through the center. This results in an additional source of dissipation and makes GeNS measurements of those resonance modes not reliable -- see Sec.~\ref{Sec:Result_interpretation}.

In this context, the procedure of balancing the disk on a lens is called \textbf{suspension} and multiple suspensions of the same sample are necessary to minimize operator-related errors. Repeating the suspension process ensures that any small inconsistencies due to possible misplacement of the sample are averaged out.
Generally, the number of suspensions required is determined empirically, based on the repeatability of the measurements. In our case, about three suspensions per sample were performed, and the mechanical loss of each resonance mode was measured around five times for a suspension. The average loss of these approximately five measurements per mode was taken as the loss result for one specific suspension. The lowest loss from all suspensions was taken as the final result for each mode. This value is presented in the following sections. Since all data for each mode was comparable within $10\%$ between suspensions, we used this value as our measurement error. Table~\ref{tab:measured samples} lists the samples measured from each ingot.

\begin{table}[t]
\caption{Nomenclature and location of each measured sample within each WW. The position refers to Fig.~\ref{wwcut}.}
    \label{tab:measured samples}
\makebox[8cm][c]{
    \begin{ruledtabular}
    \begin{tabular}{ccccc}
      \multicolumn{2}{c}{WWx} & \multicolumn{3}{c}{Position}\\
\hline

\multicolumn{2}{c}{WW00} & 01 & 02 & 03 \\
\multicolumn{2}{c}{WW01} & - & 02 & - \\
\multicolumn{2}{c}{WW02} & - & 02 & - \\
\multicolumn{2}{c}{WW03} & 01 & 02 & - \\
\multicolumn{2}{c}{WW04} & - & 02 & - \\
\hline
\multicolumn{2}{c}{WW05} & - & 02 & - \\
\multicolumn{2}{c}{WW06} & - & 02 & - \\
\hline
\multicolumn{2}{c}{WW12} & 01 & 02 & 03 \\
\multicolumn{2}{c}{WW13} & - & 02 & - \\
\multicolumn{2}{c}{WW14} & - & 02 & - \\
\end{tabular}
\end{ruledtabular}
}
\end{table}

\begin{figure*}[]
 \includegraphics[width=\textwidth]{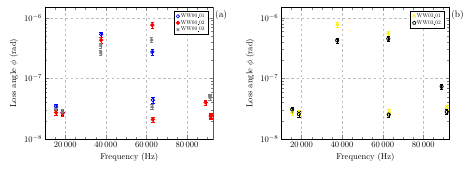}
        \caption{(a) Mechanical loss of samples from three different positions of WW00 (Ingot1) and (b) mechanical loss of two samples from different positions of WW03 (Ingot1).}
       \label{fig:combined_radial}
\end{figure*}

\subsection{Ingot1}
\label{subs:Ingot1}

Ingot1 had a usable zone approximately $588$\,mm long, from which TM01 -- TM04 were cut, along with WW00 -- WW04 and an additional small wafer, S01 -- see Fig.~\ref{fig:All_ingots_marked}. However, TM04 and S01 broke during the cutting procedure and are therefore no longer available.

\begin{figure}[t]
    \includegraphics[width=8cm]{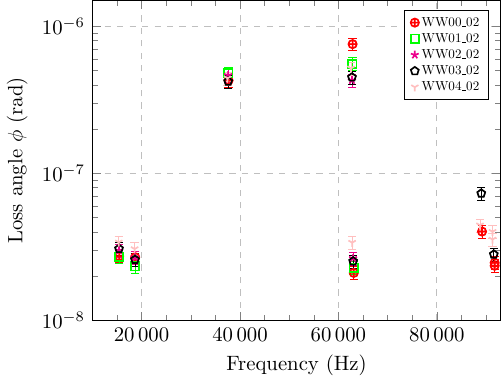}
     \caption{Mechanical loss of samples from all WWs along the length of Ingot1. All samples originate from position 02 -- see Fig.~\ref{wwcut}.}
    \label{fig:lenght_ingot1}
\end{figure}

To investigate potential radial differences within the ingot, the mechanical loss of WW$00\_01$, WW$00\_02$ and WW$00\_03$, which originate from different positions within the same WW, was measured. Additional samples, WW$03\_01$ and WW$03\_02$, from a different witness wafer, were measured as a consistency check. The results of the radial investigation are shown in Fig.~\ref{fig:combined_radial}, where Fig.~\ref{fig:combined_radial}(a) shows the mechanical loss of samples from WW$00$ and Fig.~\ref{fig:combined_radial}(b) from WW$03$. The losses measured for all five samples -- three from WW00 and two from WW03 -- show the same pattern of two modes with a higher and five modes with a lower loss level.

To investigate the mechanical loss along the length of the ingot, samples from the position $02$ of the five WWs obtained from Ingot1 were measured, and the results are shown in Fig.~\ref{fig:lenght_ingot1}.
The first five resonance modes were measured for all samples, except for the two highest-frequency modes of WW$01$ and WW$02$ which were not found.
All samples show the same pattern of five modes with a low mechanical loss and two modes with a higher loss level, as observed during the radial investigation of the loss of the ingot.

In summary, for Ingot1, the mechanical loss measured within $[15,93]$ kHz exhibits two different levels of loss: a lower level at $[2.3,7.0]\times10^{-8}$ and a higher level at $[2.7,7.8]\times10^{-7}$. A more detailed discussion of the physical mechanism responsible for the two mechanical loss levels is provided in Sec.~\ref{Sec:Result_interpretation}.

\subsection{Ingot2}
\label{Subs:Ingot2}

Ingot2 has a usable zone approximately $380$\,mm long, from which TM05 -- TM07 were cut, along with WW05 -- WW07, and two additional small wafers S02 and S03 (see Fig.~\ref{fig:All_ingots_marked}). However, WW06, WW07, TM07 and S02 broke after cutting, where some parts of WW06 were still usable. Therefore, only samples from WW05 and WW06, surrounding TM05, were measured.

\begin{figure}[t]
    \centering
    \includegraphics[width=8cm]{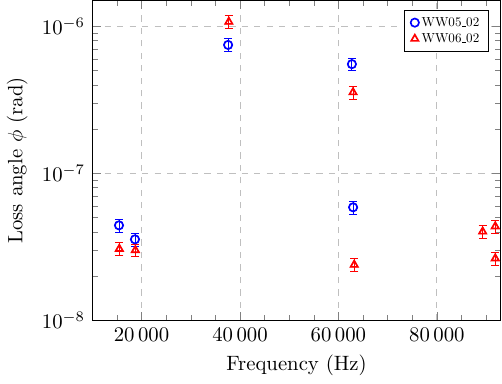}
     \caption{Mechanical loss of samples from different WWs of Ingot2. The sample originate from same position.}
    \label{fig:Results_ingot2}
\end{figure}

The results of the mechanical loss measurements are shown in Fig.~\ref{fig:Results_ingot2}, showing the same pattern of low- and high-loss modes as found for Ingot1, where
the lower mechanical loss values range between $[2.4,5.9]\times10^{-8}$, while the higher values range between $[3.5,10]\times10^{-7}$.

\subsection{Ingot4}
\label{Subs:Ingot4}

Ingot4 was the longest of the four ingots, with a usable zone of approximately $793$\,mm. A total of 7 TMs, TM11 -- TM17, were obtained, along with WW12 -- WW18, and an additional small wafer of $2$\,mm thickness -- see Fig.~\ref{fig:All_ingots_marked}.

Although Ingot1 and Ingot2 showed a high degree of homogeneity in mechanical loss along the length, radially and between ingots, it was decided to characterize a few test samples of Ingot4 for mechanical loss, as this ingot is alongside Ingot1 the most interesting one for being the longest and most complete.

Samples from all three positions within WW12 were measured. In addition, as shown in Tab.~\ref{tab:measured samples}, one sample of each WW13 and WW14 from the same position was measured to investigate possible differences along the length of the ingot.

\begin{figure}
    \centering
    \includegraphics[width=80mm]{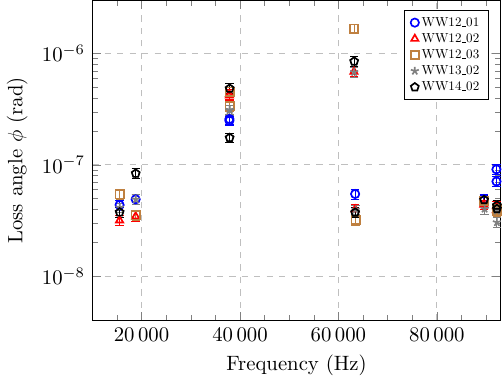}
     \caption{Mechanical loss of samples from Ingot4. This graph includes measurements from samples of different positions within WW12, as well as measurements from samples of position 02, but from different positions along the length of the ingot, i.e. different WWs.}
    \label{fig:Results_ingot4}
\end{figure}

The mechanical loss values shown in Fig.~\ref{fig:Results_ingot4} exhibit the same behavior as for Ingot1 and Ingot2, where the lower values range between $[3.0,9.0]\times10^{-8}$ while the higher values range between $[1.8,17]\times10^{-7}$.

\section{Cryogenic mechanical loss measurements}
\label{Cryogenic mechanicall loss measurements}

Since they are of particular interest to ETpathfinder, measurements at cryogenic temperatures were performed in addition to those at room temperature. 

\begin{figure}
    \includegraphics[width=80mm]{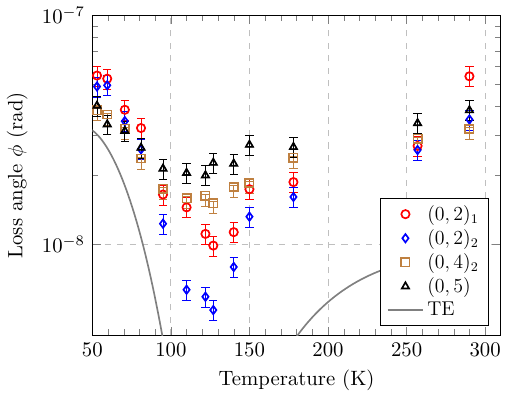}
     \caption{Temperature-dependent mechanical loss measurements, ranging from $53$\,K to $290$\,K for sample WW$12$\_$03$. Each dataset represents a different mode. The gray curve represents the calculated thermoelastic (TE) loss averaged over the four modes.}
    \label{fig:TDM_WW12_03}
\end{figure}

The cryostat uses a Gifford-McMahon cryocooler, whose two stages at 70\,K and 20\,K are thermally connected to the inner chamber of the vacuum tank. The GeNS is thermally linked to the lower-temperature stage. The sample is brought into thermal contact with the coldest stage by raising the aluminum plate of the GeNS using a voice coil actuator. Due to the heat introduced by the actuator itself, the minimum achievable temperature, during the measurements presented in this paper, was limited to 53\,K.
Temperature is monitored using a silicon diode mounted on the GeNS aluminum plate. Once the disk is cooled to its lowest achievable temperature, the plate is lowered and the disk is placed onto the GeNS lens. No drift in the resonance frequencies of the sample was observed over time, indicating that thermal contact with the lens alone is sufficient to maintain the disk at a stable temperature of 53\,K. At temperatures above $53$\,K, such as $120$\,K, thermal stability of the sample was verified by monitoring the resonance frequencies over time. The absence of drift was used as a proxy for equilibrium. Although this does not guarantee full thermal uniformity, the consistent frequency response to small temperature steps supports this assumption. However, we acknowledge that small temperature gradients or uncertainties could still be present and could contribute to an apparent smoothing of the loss minimum observed around $120$\,K.

Mechanical losses were measured for WW$12$\_$03$ in a temperature range from room temperature to $53$\,K. Heating resistors were used to achieve stable temperatures in the measurement range. At 120\,K, which is in the lowest-loss region found for WW12$\_$03 and of particular interest for one of the two possible ETpathfinder configurations~\cite{Utina_2022}, the mechanical loss was also measured for a second sample, WW$00$\_$02$.

The temperature-dependent mechanical loss measurement of WW$12$\_$03$ is shown in Fig.~\ref{fig:TDM_WW12_03}. The minima in mechanical loss for all modes at a temperature of around $120$\,K are a result of the characteristic behavior of silicon due to a zero crossing of its thermal expansion coefficient~\cite{silicon_thermal_coeff}. Continuing towards lower temperatures, the thermo-elastic loss increases again due to an increasing, although negative, magnitude of the thermal expansion coefficient~\cite{PhysRevB.92.174113}. Below approximately 20\,K, the thermal expansion coefficient approaches zero, suggesting that the losses further improve closer to absolute zero.
Figure~\ref{fig:TDM_WW12_03} only shows a selection of modes, corresponding to the lower loss level in previous figures. The higher-loss modes were excluded for better resolution. See Sec.~\ref{Sec:Result_interpretation} for a more detailed discussion of the different loss levels observed.

\begin{figure}
    \centering
    \includegraphics[width=80mm]{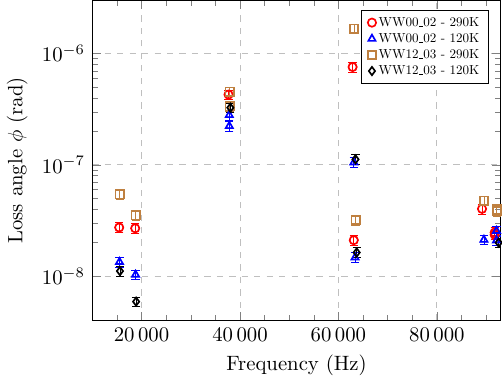}
     \caption{Comparison between room temperature and cryogenic loss angle measurements for WW$00$\textunderscore$02$ and WW$12$\textunderscore$03$..}
    \label{fig:Cryo_plot_together}
\end{figure}

The measured values obtained at $120$\,K for both samples, WW$00$\_$02$ and WW$12$\_$03$, are shown in Fig.~\ref{fig:Cryo_plot_together} together with the room temperature measurements taken on the same samples for comparison. The low-temperature mechanical loss is lower than the room-temperature loss for all modes, showing the same pattern of lower and higher losses depending on the resonance mode previously observed at room temperature.
The mechanical loss values at $120$\,K, shown in Fig.~\ref{fig:Cryo_plot_together} exhibit the same behavior as for the room temperature one, where the lower value range between $[4.6,28]\times10^{-9}$ while the higher values range between $[6.3,36]\times10^{-8}$.

\section{Result interpretation}
\label{Sec:Result_interpretation}

\begin{figure}
    \centering
    \includegraphics[width=80mm]{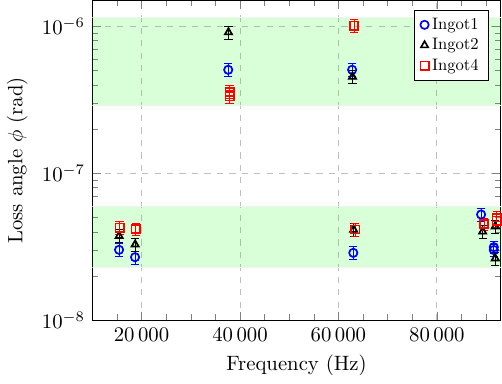}
     \caption{Average mechanical loss measurements from all the ingots. The opaque green bands in the background are intended to highlight the difference in spread and level between the lowest loss angle modes and the highest loss angle modes.}
    \label{fig:Comparison results}
\end{figure}

No significant differences were observed between the radial positions within the ingots or along the length of the ingots. Figure~\ref{fig:Comparison results} shows the mechanical loss for all resonance modes measured, averaged for each ingot presented in Sec.~\ref{Room temperature mechanical loss characterization}.

The results obtained from each ingot show that some of the modes have higher mechanical loss values accompanied by lower repeatability compared to the other modes. The green-shaded bars in Fig.~\ref{fig:Comparison results} indicate these two different loss levels.

\begin{figure*}
    \includegraphics[width=0.9\textwidth]{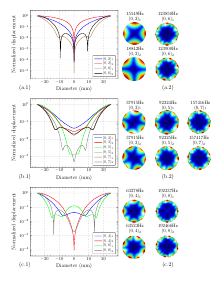}
     \caption{Figure (a.1), (b.1) and (c.1) represent the vertical displacement along the diameter of the disk for modes with respectively even, odd and multiples of 4 nodal lines. Each displacement is normalized by the maximum. Figure (a.2), (b.2) and (c.2) are graphical representations of the corresponding mode shapes. }
    \label{fig:combined_simulation}
\end{figure*}

To understand the loss behavior, a silicon disk of the same dimensions as our samples was modeled using finite-element analysis in COMSOL\footnote{\url{https://www.comsol.com}}. Figure~\ref{fig:combined_simulation} shows the surface oscillation profiles of different modes and is divided into three distinct plots. 

The crystalline structure of $\braket{100}$-oriented silicon leads to anisotropy in the horizontal plane, breaking the symmetry and the degeneracy of certain modes. 

We refer to a breaking of the mode degeneracy when the frequencies of the twin modes are no longer identical.

Generally, modes with an even number of nodal lines (Fig.~\ref{fig:combined_simulation} a.2) lose their degeneracy, while modes with an odd number of nodal lines (Fig.~\ref{fig:combined_simulation} b.2) maintain it. In fact, the same effect can be observed in isotropic materials by breaking the circular symmetry of the plane itself~\cite{Ghosh2021}. An exception to this pattern occurs with one of the twin modes that has $4$ (or an integer multiple of $4$ -- see Fig.~\ref{fig:combined_simulation} c.2) nodal lines and also exhibits this behavior. This is due to the fact that the Young’s modulus and Poisson’s ratio of $\braket{100}$-oriented silicon follow a quadrupolar pattern in the plane~\cite{Youngmodulus}.

Figure~\ref{fig:combined_simulation} a.1 shows the surface oscillation profiles of modes with an even number of nodal lines. All modes, except those with 4n nodal lines, show no oscillation in the center. This is consistent with the lower loss measured for the modes at 16 and 19\,kHz -- see Fig.~\ref{fig:Comparison results}.

Figure~\ref{fig:combined_simulation} b.1 shows the surface oscillation profiles of the modes with an odd number of nodal lines. All modes show a central oscillating area, though the amplitude for mode $(0,5)$ is lower than for the other modes. This is consistent with the higher loss measured for the mode at 38\,kHz, and the lower loss at 92\,kHz.

Figure~\ref{fig:combined_simulation} c.1 exhibits a central oscillating area for modes with 4n nodal lines, which is in agreement with this effect being due to the anisotropy caused by the quadrupolarity of the Young’s modulus and Poisson’s ratio, which drive the oscillation. In all measured samples, mode $(0,4)_1$ consistently shows higher mechanical loss values compared to its twin mode $(0,4)_2$, at 63\,kHz, making it the only exception among modes with an even number of nodal lines.

From this study, we conclude that the mechanical losses measured for the modes (0,2), $(0,4)_2$ and (0,5) is the intrinsic loss of the material, or very close to it, while that measured for modes (0,3) and $(0,4)_1$ includes losses due to external damping at the suspension point.

\begin{figure*}[t]
 \includegraphics[width=0.9\textwidth]{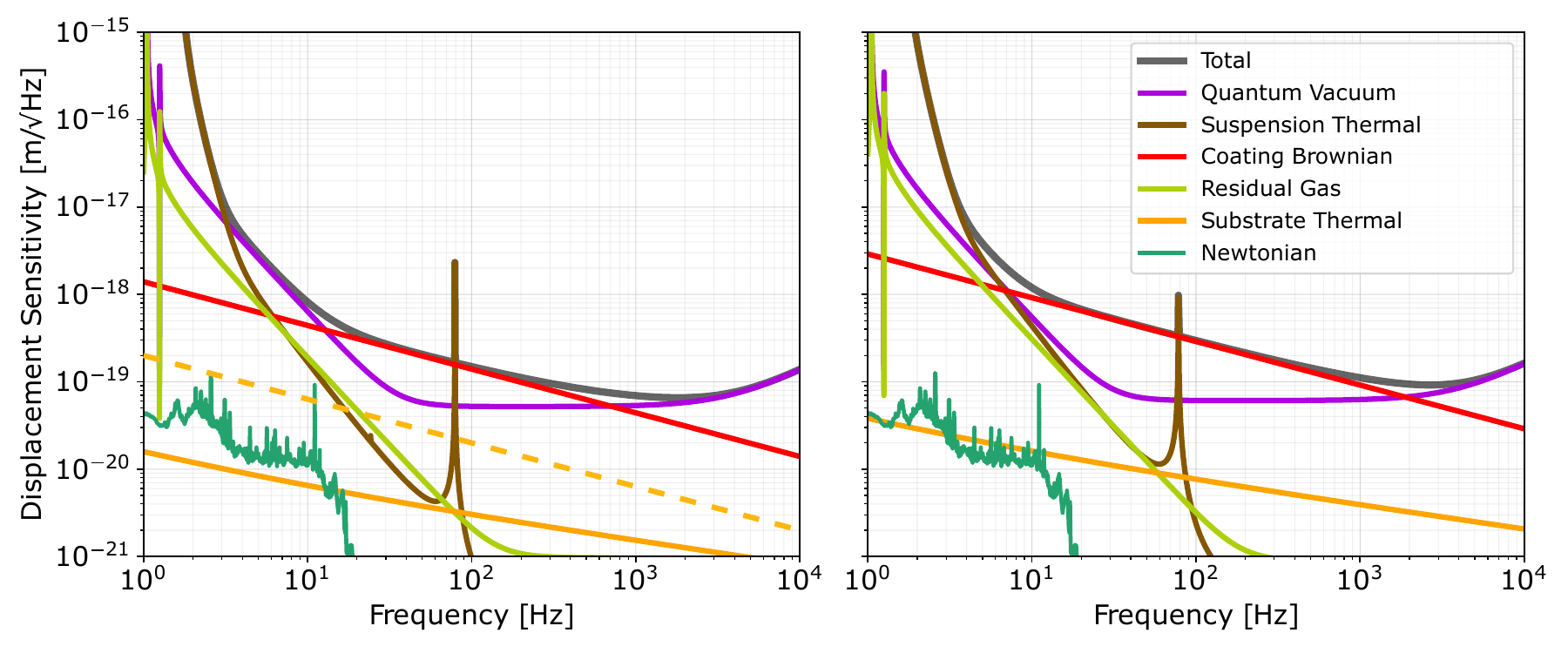}
        \caption{Left: ETpathfinderA (1550\,nm and 18\,K) and right: ETpathfinderB (2090\,nm and 123\,K). The orange lines represent thermal noise of the silicon substrates based on the mechanical loss measured (see Eq.~\ref{equ:TN}). The orange dashed line represents thermal noise of the substrate at the lowest temperature achieved in our table-top experiment, i.e. $53$\,K.}
       \label{fig:TN}
\end{figure*}

While literature suggests a frequency-dependence of the mechanical loss of silicon~\cite{LAM198141,Rodriguez}, we do not observe a frequency-dependence in loss for the lower-loss modes at room temperature. However, at 120\,K a decreasing trend in loss towards lower frequencies becomes visible -- see Fig.~\ref{fig:TDM_WW12_03} and~\ref{fig:Cryo_plot_together}. Therefore, the constant loss versus frequency observed at room temperature may be due to an increasing thermo-elastic loss component at lower frequencies, which becomes negligible at 120\,K. However, while literature suggests a constant ratio of $f/\phi$, we observe a lower loss decrease towards low frequencies of about a factor of $\approx 2$ (from $2$ to $1 \times 10^{-8}$) over a factor of $\approx 4.5$ in frequency (from 90 to 20\,kHz), see Fig.~\ref{fig:Cryo_plot_together}. This trend can be expressed by

\begin{equation}
\phi(f)\approx \phi_{\rm 0} \times \left( \frac{f}{f_{\rm 0}}\right)^{0.46},
\label{equ:phi}
\end{equation}

\noindent where $\phi_{\rm 0} \approx 1 \times 10^{-10}$ at $f_{\rm 0} = 1$\,Hz.

At 53\,K, which is the lowest temperature achieved in our setup, similar to the room-temperature measurements, no frequency-dependence of the loss is visible. This can be seen in Fig.~\ref{fig:TDM_WW12_03}), where at about 120\,K, the two lowest-frequency modes (shown in blue and red) show the lowest loss, while the relative loss level between modes changed at 53\,K. This is again consistent with an increase of thermo-elastic loss in this temperature range. Therefore, the average loss of the three lower-loss modes of sample WW12$\_03$ is taken as the final loss result for the ETpathfinder silicon, which is $\phi \approx 4.56 \times 10^{-8}$, in line with values reported in literature for crystalline silicon~\cite{R.Schnabel_2010, R.Nawrodt_2008}.

Additional mechanical loss attributed to the sample's surface can contribute to the mechanical loss as investigated by Nawrodt et al.~\cite{Nawrodt_2013}. This loss becomes more significant, the higher the surface-to-volume ratio of the sample is. For our substrate dimensions, the surface contribution is assumed to be $\leq 5$\,\%, based on Eq.
13 in~\cite{Nawrodt_2013}, and therefore does not significantly contribute to the mechanical loss measured.

\section{Substrate Thermal Noise Estimate for ET\lowercase{pathfinder}}
\label{Thermal Noise}

Figure~\ref{fig:TN} shows the envisioned sensitivity of ETpathfinder A (left -- 1550\,nm and 18\,K) and ETpathfinder B (right -- 2090\,nm and 123\,K)~\cite{Utina_2022}. Initially, coatings made of SiO$_2$ and Ta$_2$O$_5$ are planned to be used. The substrate thermal noise level is shown by the orange line in both graphs, and additionally by the orange, dashed line for ETpathfinder A.

Substrate thermal noise can be expressed by~\cite{BONDU1998227}

\begin{equation}
\label{equ:TN}
S_{\rm sub}= \sqrt{\frac{2k_{\rm B}T(1-\sigma^2)}{\sqrt{\pi^3}fYw_0}\phi(f)}\,.
\end{equation}

\noindent For ETpathfinder B, the frequency-dependent mechanical loss $\phi (f)$ described by Eq.~\ref{equ:phi} was used.
For ETpathfinder A, two scenarios are shown:

\begin{enumerate}
    \item For the dashed, orange (top) line, a frequency-independent mechanical loss of $\phi = 4.56 \times 10^{-8}$ was used -- see end of Sec.~\ref{Sec:Result_interpretation} for details. This is the loss measured at 53\,K, which was the lowest, and therefore closest to 18\,K, temperature at which we were able to measure in our setup. This substrate loss, likely over-estimates substrate thermal noise, as the mechanical loss is expected to decrease between 53 and 18\,K to the loss level at 120\,K or below~\cite{PhysRevResearch.4.043043, Nawrodt_2013}.
    \item For the solid orange (bottom) line, which is the substrate thermal noise level contributing to the total noise (gray curve), the same frequency dependent loss, measured at approximately 120\,K, was used as for ETpathfinder B. This is likely the better loss estimate at 18\,K, but might still over-estimate substrate thermal noise. 
\end{enumerate}

In both cases the thermoelastic noise contribution has been considered negligible. Thermo-refractive noise has not been included, however we recognize that it may affect the total noise budget~\cite{Meyer_2022}. In~\cite{Nawrodt_2013}, Nawrodt et al. investigated the surface loss contribution to the substrate thermal noise, which is calculated equivalent to coating thermal noise with a loss-thickness product of $\phi \times d = 5\times 10^{-13}$\,m. This additional loss source was also considered for the calculation of the substrate thermal noise levels shown in Fig.~\ref{fig:TN}.

For all scenarios discussed, the estimated substrate thermal noise level is significantly lower than coating thermal noise (red), which dominates the total ETpathfinder noise (grey) over a wide frequency range. In the event of a significant coating thermal noise reduction, e.g. due to the use of lower-noise coatings, substrate thermal noise will only become significant for upper limit indicated by the dashed, orange line for ETpathfinder A. However, it is unlikely that this upper limit will be reached.

\section{Conclusions}
\label{Conclusions}

ETpathfinder is a prototype for the next generation of gravitational-wave detectors such as the Einstein Telescope, operating at low temperature with silicon as a test mass material~\cite{Utina_2022}. This study provides a comprehensive analysis of the mechanical loss characteristics of silicon ingots, from which the ETpathfinder test masses will be obtained, highlighting the potential of crystalline silicon as a low-loss substrate material suitable for use in cryogenic environments.

Float zone silicon from three different ingots was characterized for its mechanical loss in a temperature range between 53\,K, which was the lowest temperature our setup achieved, and room temperature. No significant variation in mechanical loss was observed between the ingots or between the radial or longitudinal positions within individual ingots. This consistency supports the overall homogeneity and high quality of the material, which is crucial to ensure the reliability of the ETpathfinder test masses.

One of the findings of our study was the possible influence of the anisotropic crystalline structure of silicon on the mechanical loss results, particularly in modes with odd numbers of nodal lines, where higher dissipation is a consequence of symmetry breaking. We believe, this phenomenon introduces a secondary source of loss that must be considered when using a GeNS for mechanical loss measurements.

Based on the mechanical loss found for modes, which are not affected by additional damping, we made an estimate of the upper limit of substrate thermal noise expected for the two different scenarios of ETpathfinder operating at 18\,K and 123\,K. We show that the expected substrate thermal noise level is at least a factor of 10 lower than coating thermal noise in the initial phase of ETpathfinder. Future studies on crystalline silicon could expand the temperature range for cryogenic testing and explore methods to minimize the additional source of dissipation.

\section{Acknowledgements}

This work is supported by ERC grant MIRRORS, Project No. 101040572, funded by the European Union. Views and opinions expressed are however those of the authors only and do not necessarily reflect those of the European Union or the European Research Council Executive Agency. Neither the European Union nor the granting authority can be held responsible for them.

We acknowledge support by ETpathfinder (Interreg Vlaanderen-Nederland), E-TEST (Interreg Euregio Meuse-Rhine), the Dutch Research Council (NWO) Project No. VI.Vidi.203.062 and OCENW.KLEIN.560 and the Province of Limburg.

This paper has LIGO Document number LIGO-P2500164.

\bibliographystyle{elsart-num.bst}
\bibliography{Mechanical_Characterisation_of_Silicon_for_the_ETpathfinder_Test_Masses}

\begin{thebibliography}{10}
\expandafter\ifx\csname url\endcsname\relax
  \def\url#1{\texttt{#1}}\fi
\expandafter\ifx\csname urlprefix\endcsname\relax\def\urlprefix{URL }\fi

\bibitem{PhysRevLett.116.061102}
B.~P. Abbott, R.~Abbott, T.~D. Abbott, M.~R. Abernathy, F.~Acernese, K.~Ackley, C.~Adams, T.~Adams, P.~Addesso, Adhikari, et~al., Observation of gravitational waves from a binary black hole merger, Phys. Rev. Lett. 116 (2016) 061102.
\newline\urlprefix\url{https://link.aps.org/doi/10.1103/PhysRevLett.116.061102}

\bibitem{GW170817}
B.~P. Abbott, R.~Abbott, T.~D. Abbott, F.~Acernese, K.~Ackley, C.~Adams, T.~Adams, P.~Addesso, R.~X. Adhikari, Adya, et~al., {GW170817}: Observation of gravitational waves from a binary neutron star inspiral, Phys. Rev. Lett. 119 (2017) 161101.
\newline\urlprefix\url{https://link.aps.org/doi/10.1103/PhysRevLett.119.161101}

\bibitem{Multi-messenger_Observations}
B.~P. Abbott, R.~Abbott, T.~D. Abbott, F.~Acernese, K.~Ackley, C.~Adams, T.~Adams, P.~Addesso, R.~X. Adhikari, Adya, et~al., Multi-messenger observations of a binary neutron star merger*, The Astrophysical Journal Letters 848~(2) (2017) L12.
\newline\urlprefix\url{https://dx.doi.org/10.3847/2041-8213/aa91c9}

\bibitem{universe2030022}
J.~L. Cervantes-Cota, S.~Galindo-Uribarri, G.~F. Smoot, A brief history of gravitational waves, Universe 2~(3).
\newline\urlprefix\url{https://www.mdpi.com/2218-1997/2/3/22}

\bibitem{Aasi_2015}
{The {LIGO} Scientific Collaboration}, J.~Aasi, B.~P. Abbott, R.~Abbott, T.~Abbott, M.~R. Abernathy, K.~Ackley, C.~Adams, T.~Adams, P.~Addesso, R.~X. Adhikari, Advanced {LIGO}, Classical and Quantum Gravity 32~(7) (2015) 074001.
\newline\urlprefix\url{https://dx.doi.org/10.1088/0264-9381/32/7/074001}

\bibitem{Acernese_2015}
F.~Acernese, M.~Agathos, K.~Agatsuma, D.~Aisa, N.~Allemandou, A.~Allocca, J.~Amarni, P.~Astone, G.~Balestri, G.~Ballardin, F.~Barone, J.-P. Baronick, Advanced virgo: a second-generation interferometric gravitational wave detector, Classical and Quantum Gravity 32~(2) (2014) 024001.
\newline\urlprefix\url{https://dx.doi.org/10.1088/0264-9381/32/2/024001}

\bibitem{KAGRA_run_join}
B.~P. Abbott, R.~Abbott, T.~D. Abbott, S.~Abraham, F.~Acernese, K.~Ackley, C.~Adams, V.~B. Adya, C.~Affeldt, M.~Agathos, et~al., Prospects for observing and localizing gravitational-wave transients with advanced {LIGO}, advanced virgo and {KAGRA}, Living Reviews in Relativity.
\newline\urlprefix\url{https://doi.org/10.1007/s41114-020-00026-9}

\bibitem{ETdesignstudy}
E.~T. steering committee, et~al., Einstein telescope: Science case, design study and feasibility report (2020).

\bibitem{Saulson}
P.~R. Saulson, Thermal noise in mechanical experiments, Phys. Rev. D 42 (1990) 2437--2445.
\newline\urlprefix\url{https://link.aps.org/doi/10.1103/PhysRevD.42.2437}

\bibitem{10.1063/1.1394183}
S.~D. Penn, G.~M. Harry, A.~M. Gretarsson, S.~E. Kittelberger, P.~R. Saulson, J.~J. Schiller, J.~R. Smith, S.~O. Swords, {High quality factor measured in fused silica}, Review of Scientific Instruments 72~(9) (2001) 3670--3673.
\newline\urlprefix\url{https://doi.org/10.1063/1.1394183}

\bibitem{schroeter2007mechanical}
A.~Schroeter, R.~Nawrodt, R.~Schnabel, S.~Reid, I.~Martin, S.~Rowan, C.~Schwarz, T.~Koettig, R.~Neubert, M.~Thürk, W.~Vodel, A.~Tünnermann, K.~Danzmann, P.~Seidel, On the mechanical quality factors of cryogenic test masses from fused silica and crystalline quartz (2007).

\bibitem{10.1063/1.1721138}
J.~W. Marx, J.~M. Sivertsen, Temperature dependence of the elastic moduli and internal friction of silica and glass, Journal of Applied Physics 24~(1) (1953) 81--87.
\newline\urlprefix\url{https://doi.org/10.1063/1.1721138}

\bibitem{1954JAP.25.402F}
M.~E. {Fine}, H.~{Van Duyne}, N.~T. {Kenney}, {Low-Temperature Internal Friction and Elasticity Effects in Vitreous Silica}, Journal of Applied Physics 25~(3) (1954) 402--405.

\bibitem{R_Nawrodt_2008}
R.~Nawrodt, A.~Zimmer, T.~Koettig, C.~Schwarz, D.~Heinert, M.~Hudl, R.~Neubert, M.~Thürk, S.~Nietzsche, W.~Vodel, P.~Seidel, A.~Tünnermann, High mechanical q-factor measurements on silicon bulk samples, Journal of Physics: Conference Series 122~(1) (2008) 012008.
\newline\urlprefix\url{https://dx.doi.org/10.1088/1742-6596/122/1/012008}

\bibitem{UCHIYAMA19995}
T.~Uchiyama, T.~Tomaru, M.~Tobar, D.~Tatsumi, S.~Miyoki, M.~Ohashi, K.~Kuroda, T.~Suzuki, N.~Sato, T.~Haruyama, A.~Yamamoto, T.~Shintomi, Mechanical quality factor of a cryogenic sapphire test mass for gravitational wave detectors, Physics Letters A 261~(1) (1999) 5--11.
\newline\urlprefix\url{https://www.sciencedirect.com/science/article/pii/S0375960199005630}

\bibitem{10.1117/12.459019}
S.~Rowan, R.~L. Byer, M.~M. Fejer, R.~K. Route, G.~Cagnoli, D.~R. Crooks, J.~Hough, P.~H. Sneddon, W.~Winkler, {Test mass materials for a new generation of gravitational wave detectors}, in: P.~Saulson, A.~M. Cruise (Eds.), Gravitational-Wave Detection, Vol. 4856, International Society for Optics and Photonics, SPIE, 2003, pp. 292 -- 297.
\newline\urlprefix\url{https://doi.org/10.1117/12.459019}

\bibitem{Utina_2022}
A.~Utina, A.~Amato, J.~Arends, C.~Arina, M.~de~Baar, M.~Baars, P.~Baer, N.~van Bakel, W.~Beaumont, A.~Bertolini, et~al., Etpathfinder: a cryogenic testbed for interferometric gravitational-wave detectors, Classical and Quantum Gravity 39~(21) (2022) 215008.
\newline\urlprefix\url{https://dx.doi.org/10.1088/1361-6382/ac8fdb}

\bibitem{Silenzi_2024}
L.~Silenzi, F.~Fabrizi, M.~Granata, L.~Mereni, M.~Montani, F.~Piergiovanni, A.~Trapananti, F.~Travasso, G.~Cagnoli, Towards the solution of coating loss measurements using thermoelastic-dominated substrates, Classical and Quantum Gravity 41~(23) (2024) 235017.
\newline\urlprefix\url{https://dx.doi.org/10.1088/1361-6382/ad8543}

\bibitem{theses_murray}
P.~G. Murray, Measurement of the mechanical loss of test mass materials for advanced gravitational wave detectors, phD (2008).
\newline\urlprefix\url{https://theses.gla.ac.uk/565/}

\bibitem{Cesarini_2009}
E.~Cesarini, M.~Lorenzini, E.~Campagna, F.~Martelli, F.~Piergiovanni, F.~Vetrano, G.~Losurdo, G.~Cagnoli, A "gentle" nodal suspension for measurements of the acoustic attenuation in materials, Review of Scientific Instruments 80~(5) (2009) 053904.
\newline\urlprefix\url{https://doi.org/10.1063/1.3124800}

\bibitem{silicon_thermal_coeff}
J.~Rodriguez, S.~A. Chandorkar, G.~M. Glaze, D.~D. Gerrard, Y.~Chen, D.~B. Heinz, I.~B. Flader, T.~W. Kenny, Direct detection of anchor damping in mems tuning fork resonators, Journal of Microelectromechanical Systems 27~(5) (2018) 800--809.

\bibitem{PhysRevB.92.174113}
T.~Middelmann, A.~Walkov, G.~Bartl, R.~Sch\"odel, Thermal expansion coefficient of single-crystal silicon from 7 k to 293 k, Phys. Rev. B 92 (2015) 174113.
\newline\urlprefix\url{https://link.aps.org/doi/10.1103/PhysRevB.92.174113}

\bibitem{Ghosh2021}
A.~Ghosh, A.~DasGupta, Vibration analysis of irregular-shaped plates on simple supports, Proceedings of the Royal Society A: Mathematical, Physical and Engineering Sciences 477~(2252) (2021) 20210184.
\newline\urlprefix\url{https://royalsocietypublishing.org/doi/abs/10.1098/rspa.2021.0184}

\bibitem{Youngmodulus}
M.~A. Hopcroft, W.~D. Nix, T.~W. Kenny, What is the young's modulus of silicon?, Journal of Microelectromechanical Systems 19~(2) (2010) 229--238.

\bibitem{LAM198141}
C.~Lam, D.~Douglass, Internal friction measurements in boron-doped single-crystal silicon, Physics Letters A 85~(1) (1981) 41--42.
\newline\urlprefix\url{https://www.sciencedirect.com/science/article/pii/0375960181906356}

\bibitem{Rodriguez}
W.~C. Rodriguez~J., Chandorkar~S.A., et~al., Direct detection of akhiezer damping in a silicon mems resonator, Sci Rep 9~(2244).
\newline\urlprefix\url{https://doi.org/10.1038/s41598-019-38847-6}

\bibitem{R.Schnabel_2010}
R.~Schnabel, M.~Britzger, F.~Brückner, O.~Burmeister, K.~Danzmann, J.~Duck, T.~Eberle, D.~Friedrich, H.~Luck, M.~Mehmet, R.~Nawrodt, S.~Steinlechner, B.~Willke, Building blocks for future detectors: Silicon test masses and 1550 nm laser light, Journal of Physics: Conference Series 228~(1) (2010) 012029.
\newline\urlprefix\url{https://dx.doi.org/10.1088/1742-6596/228/1/012029}

\bibitem{R.Nawrodt_2008}
R.~Nawrodt, A.~Zimmer, T.~Koettig, C.~Schwarz, D.~Heinert, M.~Hudl, R.~Neubert, M.~Thürk, S.~Nietzsche, W.~Vodel, P.~Seidel, A.~Tünnermann, High mechanical q-factor measurements on silicon bulk samples, Journal of Physics: Conference Series 122~(1) (2008) 012008.
\newline\urlprefix\url{https://dx.doi.org/10.1088/1742-6596/122/1/012008}

\bibitem{Nawrodt_2013}
R.~Nawrodt, C.~Schwarz, S.~Kroker, I.~W. Martin, R.~Bassiri, F.~Brückner, L.~Cunningham, G.~D. Hammond, D.~Heinert, J.~Hough, T.~Käsebier, E.-B. Kley, R.~Neubert, S.~Reid, S.~Rowan, P.~Seidel, A.~Tünnermann, Investigation of mechanical losses of thin silicon flexures at low temperatures, Classical and Quantum Gravity 30~(11) (2013) 115008.
\newline\urlprefix\url{https://dx.doi.org/10.1088/0264-9381/30/11/115008}

\bibitem{BONDU1998227}
F.~Bondu, P.~Hello, J.-Y. Vinet, Thermal noise in mirrors of interferometric gravitational wave antennas, Physics Letters A 246~(3) (1998) 227--236.
\newline\urlprefix\url{https://www.sciencedirect.com/science/article/pii/S0375960198004502}

\bibitem{PhysRevResearch.4.043043}
F.~M. Kiessling, P.~G. Murray, M.~Kinley-Hanlon, I.~Buchovska, T.~K. Ervik, V.~Graham, J.~Hough, R.~Johnston, M.~Pietsch, S.~Rowan, R.~Schnabel, S.~C. Tait, J.~Steinlechner, I.~W. Martin, Quasi-monocrystalline silicon for low-noise end mirrors in cryogenic gravitational-wave detectors, Phys. Rev. Res. 4 (2022) 043043.
\newline\urlprefix\url{https://link.aps.org/doi/10.1103/PhysRevResearch.4.043043}

\bibitem{Meyer_2022}
J.~Meyer, W.~Dickmann, S.~Kroker, M.~Gaedtke, J.~Dickmann, Thermally induced refractive index fluctuations in transmissive optical components and their influence on the sensitivity of einstein telescope, Classical and Quantum Gravity 39~(13) (2022) 135001.
\newline\urlprefix\url{https://dx.doi.org/10.1088/1361-6382/ac6e21}

\end{thebibliography}

\end{document}